\documentclass[aps,prl,twocolumn,superscriptaddress,groupedaddress]{revtex4}

\usepackage{graphicx}
\usepackage{dcolumn}
\usepackage{bm}
\usepackage{hyperref}
\usepackage[mathlines]{lineno}
\usepackage{amsfonts}
\usepackage{amssymb}

\usepackage{mathtools}

\newlength{\thelinewidth}
\thelinewidth=\textwidth

\begin{document}

\title{Axion search with a quantum-limited ferromagnetic haloscope}

\author{N.~Crescini}
	\email{nicolo.crescini@phd.unipd.it}
	\affiliation{INFN-Laboratori Nazionali di Legnaro, Viale dell'Universit\`a 2, 35020 Legnaro (PD), Italy}
	\affiliation{Dipartimento di Fisica e Astronomia, Via Marzolo 8, 35131 Padova, Italy}
		
\author{D.~Alesini}
	\affiliation{INFN-Laboratori Nazionali di Frascati, Via Enrico Fermi 40, 00044 Roma, Italy}
	
\author{C.~Braggio}
	\affiliation{Dipartimento di Fisica e Astronomia, Via Marzolo 8, 35131 Padova, Italy}
	\affiliation{INFN-Sezione di Padova, Via Marzolo 8, 35131 Padova, Italy}
	
\author{G.~Carugno}
	\affiliation{Dipartimento di Fisica e Astronomia, Via Marzolo 8, 35131 Padova, Italy}
	\affiliation{INFN-Sezione di Padova, Via Marzolo 8, 35131 Padova, Italy}
	
\author{D.~D'Agostino}
	\affiliation{INFN-Sezione di Napoli, Via Cinthia, 80126 Napoli, Italy and Dipartimento di Fisica, Via Giovanni Paolo II 132, 84084 Fisciano (SA), Italy}
	
\author{D.~Di Gioacchino}
	\affiliation{INFN-Laboratori Nazionali di Frascati, Via Enrico Fermi 40, 00044 Roma, Italy}
	
\author{P. Falferi}
	\affiliation{IFN-CNR, Fondazione Bruno Kessler, and INFN-TIFPA, Via alla Cascata 56, 38123 Povo (TN), Italy}	
	
\author{U.~Gambardella}
	\affiliation{INFN-Sezione di Napoli, Via Cinthia, 80126 Napoli, Italy and Dipartimento di Fisica, Via Giovanni Paolo II 132, 84084 Fisciano (SA), Italy}
	
\author{C.~Gatti}
	\affiliation{INFN-Laboratori Nazionali di Frascati, Via Enrico Fermi 40, 00044 Roma, Italy}
	
\author{G.~Iannone}
	\affiliation{INFN-Sezione di Napoli, Via Cinthia, 80126 Napoli, Italy and Dipartimento di Fisica, Via Giovanni Paolo II 132, 84084 Fisciano (SA), Italy}
	
\author{C.~Ligi}
	\affiliation{INFN-Laboratori Nazionali di Frascati, Via Enrico Fermi 40, 00044 Roma, Italy}
	
\author{A.~Lombardi}
	\affiliation{INFN-Laboratori Nazionali di Legnaro, Viale dell'Universit\`a 2, 35020 Legnaro (PD), Italy}
	
\author{A.~Ortolan}
	\affiliation{INFN-Laboratori Nazionali di Legnaro, Viale dell'Universit\`a 2, 35020 Legnaro (PD), Italy}
	
\author{R.~Pengo}
	\affiliation{INFN-Laboratori Nazionali di Legnaro, Viale dell'Universit\`a 2, 35020 Legnaro (PD), Italy}
	
\author{G.~Ruoso}
	\email{ruoso@lnl.infn.it}
	\affiliation{INFN-Laboratori Nazionali di Legnaro, Viale dell'Universit\`a 2, 35020 Legnaro (PD), Italy}
	
\author{L.~Taffarello}
	\affiliation{INFN-Sezione di Padova, Via Marzolo 8, , 35131 Padova, Italy}

\collaboration{QUAX Collaboration}

\date{\today}

\begin{abstract}
A ferromagnetic axion haloscope searches for Dark Matter in the form of axions by exploiting their interaction with electronic spins.
It is composed of an axion-to-electromagnetic field transducer coupled to a sensitive rf detector.
The former is a photon-magnon hybrid system, and the latter is based on a quantum-limited Josephson parametric amplifier.
The hybrid system consists of ten 2.1\,mm diameter YIG spheres coupled to a single microwave cavity mode by means of a static magnetic field. 
Our setup is the most sensitive rf spin-magnetometer ever realized. The minimum detectable field is $5.5\times10^{-19}\,$T with 9\,h integration time, corresponding to a limit on the axion-electron coupling constant $g_{aee}\le1.7\times10^{-11}$ at 95\% CL. 
The scientific run of our haloscope resulted in the best limit on DM-axions to electron coupling constant in a frequency span of about 120\,MHz, corresponding to the axion mass range $42.4$-$43.1\,\mu$eV. This is also the first apparatus to perform an axion mass scanning by changing the static magnetic field.
\end{abstract}

\maketitle

The axion is a beyond the Standard Model (BSM) hypothetical particle, first introduced in the seventies as a consequence of the strong CP problem of quantum chromodynamics (QCD) \cite{pq,weinberg1978new,wilczek1978problem}. Present experimental efforts are directed towards ``invisible'' axions, described by the KSVZ \cite{PhysRevLett.43.103,SHIFMAN1980493} and DFSZ \cite{DINE1983137,Zhitnitsky:1980he} models, which are extremely light and weakly coupled to the Standard Model particles.
Axions can be produced in the early Universe by different mechanisms \cite{PRESKILL1983127,Sikivie2008,Duffy_2009,MARSH20161}, and may be the main constituents of galactic Dark Matter (DM) halos.
Astrophysical and cosmological constraints \cite{RAFFELT19901,turner1990windows}, as well as lattice QCD calculations of the DM density \cite{borsanyi2016calculation,berkowitz2015lattice}, provide a preferred axion mass window around tens of $\mu$eV.

Non-baryonic DM is where cosmology meets particle physics, and axions are among the most interesting and challenging BSM particles to detect. Their experimental search can be carried out with Earth-based instruments immersed in the Milky Way's halo, which are therefore called ``haloscopes'' \cite{PhysRevLett.51.1415}. Nowadays, haloscopes rely on the inverse Primakoff effect to detect axion-induced excess photons inside a microwave cavity in a static magnetic field. Primakoff haloscopes allowed to exclude axions with masses $m_a$ between 1.91 and 3.69\,\,$\mu$eV \cite{PhysRevLett.120.151301,PhysRevLett.104.041301,admx_arxiv}, and, together with helioscopes \cite{Anastassopoulos2017}, are the only experiments which reached the QCD-axion parameter space. 
The last years saw a flourishing of new ideas to search for axions and axion-like-particles (ALPs) \cite{axion_searches,ringwald,redondo,caldwell2017dielectric,PhysRevX.4.021030,MCALLISTER201767,PhysRevLett.122.121802,PhysRevD.92.092002,EHRET2010149,PhysRevLett.122.191302,PhysRevLett.120.161801,CRESCINI2017677,pdg}.
Among these, the QUAX experiment \cite{BARBIERI2017135,quaxepjc} searches for DM axions through their coupling with the spin of the electron. This experiment aims to implement the idea of Ref.\,\cite{BARBIERI1989357} as follows.

The axion-electron interaction is described by the Lagrangian
\begin{equation}
{\cal L}_{ae} = \frac{g_{aee}}{2m_e}\partial_\mu a \big(\bar{\psi}_e\gamma^\mu\gamma_5 \psi_e \big),
\label{eq:laee}
\end{equation}
where $g_{aee}$ is the axion-electron interaction constant, $a$ is the axion field, $\psi_e$ and $m_e$ are the electron wavefunction and mass, and $\gamma_\mu$ and $\gamma_5$ are Dirac matrices. This vertex describes an axion-induced flip of an electron spin, which then decays back to the ground state emitting a photon. 
Since $v_a$, the relative speed between Earth and the DM halo, is small, we may use the non-relativistic limit of Euler-Lagrange equations and recast the interaction term
\begin{equation}
{\cal L}_{ae} \simeq -2\mu_B \boldsymbol{\sigma} \cdot \Big( \frac{g_{aee}}{2e} \Big)\nabla a \equiv -2\mu_B \boldsymbol{\sigma} \cdot \mathbf{B}_a.
\label{eq:}
\end{equation}
Here $-2\mu_B\boldsymbol{\sigma}$ and $e$ are the spin and charge of the electron, $\mu_B$ is the Bohr magneton, and $\mathbf{B}_a$ is defined as the axion effective magnetic field.
As $\nabla a \propto v_a$ \cite{BARBIERI1989357}, the non-zero value of $v_a$ results in $\mathbf{B}_a\neq0$.
	
If accounting for the whole DM, the numeric axion density is $n_a\simeq8\times10^{12}\,(42\,\mu\mathrm{eV}/m_a)\,\mathrm{cm}^{-3}$. 
For $v_a\simeq10^{-3}c$, where $c$ is the speed of light, the de Broglie wavelength and coherence time of the galactic axion field are $\lambda_{\nabla a}=25\,(42\,\mu\mathrm{eV}/m_a)$\,m, and $\tau_{\nabla a}=85\,(m_a/42\,\mu\mathrm{eV})\,\mu$s \cite{BARBIERI2017135,quaxepjc}.
The effective field frequency is proportional to the axion mass, $\omega_a/2\pi= 10\,(m_a/42\,\mu\mathrm{eV})\,\mathrm{GHz}$, while its amplitude depends on the properties of the DM halo and of the axion model,
\begin{equation}
B_a = \frac{g_{aee}}{2e}\sqrt{ \frac{n_a \hbar}{m_a c}}  m_a v_a \simeq 4\times10^{-23}\, \Big( \frac{m_a}{42\,\mu\mathrm{eV}} \Big)\,\mathrm{T},
\label{eq:ba}
\end{equation}
where $\hbar$ is the reduced Planck constant.
These features allows for the driving of a coherent interaction between $\mathbf{B}_a$ and the homogeneous magnetization of a macroscopic sample. The sample is immersed in a static magnetic field $B_0$ to couple the axion field to the Kittel mode of uniform precession of the magnetization. 
The interaction yields a conversion rate of axions to magnons which can be measured by searching for oscillations in the sample's magnetization. 
Due to the angle between $\mathbf{B}_0$ and $\mathbf{B}_a$, the resulting signal undergoes a full daily modulation \cite{Knirck_2018}.
The maximum axion-deposited power is related to Eq.\,(\ref{eq:ba}) and to the characteristics of the receiver, namely number of spins $N_s$ and system relaxation time $\tau_s$
\begin{equation}
P_a=\gamma_e \mu_B N_s \omega_a B_a^2 \tau_s,
\label{eq:2_ax_dep}
\end{equation}
where $\gamma_e$ is the electron gyromagnetic ratio.

To detect this signal we devised a suitable receiver. As it measures the magnetization of a sample, it is configured as a spin-magnetometer  used as an axion haloscope. The device consists of an axion field transducer and of an rf detection chain.

At high frequencies and in free field, the electron spin resonance linewidth is dominated by radiation damping, which limits $\tau_s$ \cite{PhysRev.95.8,doi:10.1063/1.1722859,AUGUSTINE2002111}. To avoid this issue, the material is placed in a microwave cavity.
If the frequency of the Kittel mode $\omega_m=\gamma_e B_0$ is close to the cavity mode frequency $\omega_c$, the two resonances hybridize and the single mode splits into two, following an anticrossing curve \cite{PhysRevLett.113.083603,PhysRevLett.113.156401}. The $B_0$-dependent hybrid modes frequencies are $\omega_1$ and $\omega_2$ and the cavity-material coupling is $g_{cm}=\min(\omega_2-\omega_1$). If $g_{cm}$ is larger than the hybrid mode linewidths $\gamma_{1,2}\simeq(\gamma_c+\gamma_m)/2$, where $\gamma_c$ and $\gamma_m$ are the cavity and material dissipations, the system is in the strong coupling regime. 
To increase $P_a$, $N_s$ and $\tau_s$ must be large, so a suitable sample has a high spin density and a narrow linewidth. The best material identified so far is Yttrium Iron Garnet (YIG), with roughly $2\times10^{22}\,\mathrm{spins/cm^3}$ and 1\,MHz linewidth \cite{PhysRev.110.1311}.

In the apparatus that we operated at the Laboratori Nazionali di Legnaro of INFN, the TM110 mode of a cylindrical copper cavity is coupled to ten 2.1\,mm-diameter spheres of YIG. The spherical shape is needed to avoid geometrical demagnetization. We devised an on-site grinding and polishing procedure to obtain narrow linewidth spherical samples starting from large single-crystals of YIG.
The spheres are placed on the axis of the cavity, where the rf magnetic field is uniform.

Several room temperature tests were performed to design the YIG holder: a 4\,mm inner diameter fused silica pipe, containing 10 stacked PTFE cups, each one large enough to host a free rotating YIG.
Free rotation permits the spheres' easy axis self-alignment to the external magnetic field, while a separation of 3\,mm prevents sphere-sphere interaction.
The pipe is filled with 1\,bar of helium and anchored to the cavity for thermalization.
The cavity and pipe are placed inside the internal vacuum chamber (IVC) of a dilution refrigerator, with a base temperature around 90\,mK.
Outside the IVC, in a liquid helium bath, a superconducting magnet provides the static field with an inhomogeneity below 7\,ppm over all the spheres.

	\begin{figure}[h!]
	\centering
	\includegraphics[width=.27\textwidth]{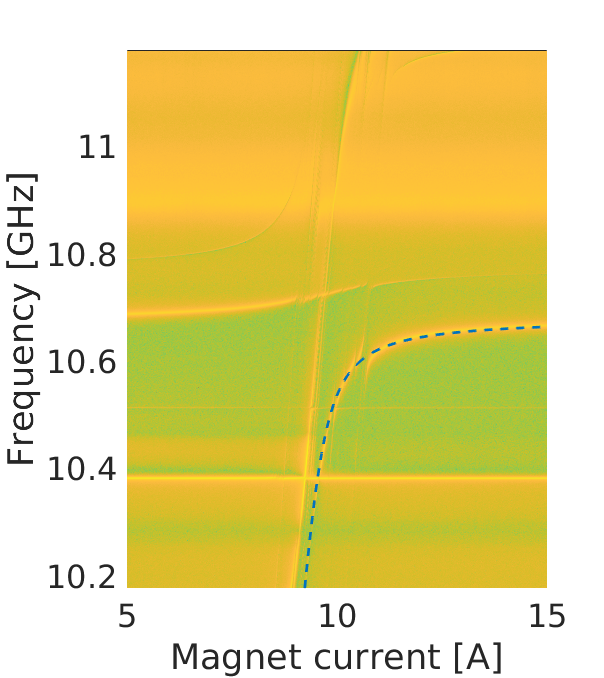}
	\includegraphics[width=.195\textwidth]{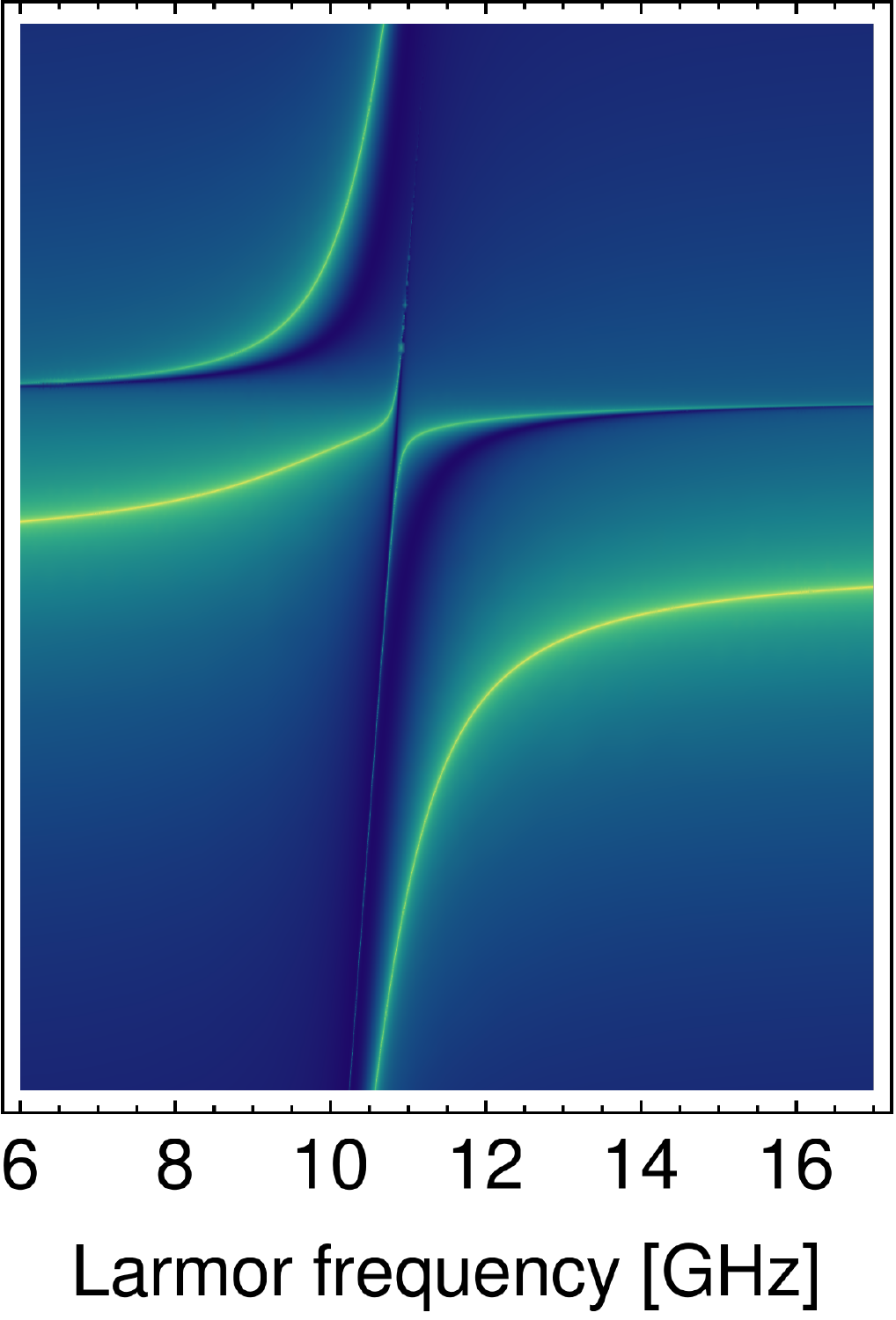}
	\caption{Measured (left) and modeled (right) transmission functions of the HS. The right plot is the function $f_{cdmn}(\omega,\omega_m)$, based on the second quantization of coupled harmonic oscillators, while the left one is a SO-to-Readout (see Fig.\,\ref{fig:app}) transmission measurement with the JPA off, performed at 90\,mK. Color scales are in arbitrary units (brighter colors corresponds to higher amplitudes). The dashed line in the left plot identifies the hybrid mode frequencies $\omega_1$, where we performed measurements.}
	\label{fig:tuning}
	\end{figure}
The resulting hybrid system (HS) has been studied by collecting a $B_0$ vs frequency transmission plot, reported in Fig.\,\ref{fig:tuning} (left).
The measured plot is not a usual anticrossing curve. 
In our system the cavity frequency $\omega_c/2\pi=10.7\,$GHz and the expected coupling is of the order of 600\,MHz, thus $\omega_2$ gets close and couples to a higher order mode of the cavity. This hybrid mode further splits into others, making the two oscillators description unsuitable. Other disturbances are related to residual sphere-sphere interaction and to non-identical spheres. 
To model the HS, we write an hamiltonian based on two cavity modes, $c$ and $d$, and two magnetic modes, $m$ and $n$
\begin{equation}
{\cal H}_{cdmn} =
	\begin{pmatrix}
	\omega_c -\frac{i\gamma_c}{2} 	&	0	&	g_{cm}		&	g_{cn}		\\
	0			&	\omega_d-\frac{i\gamma_d}{2}	&	g_{dm}	&	g_{dn}	\\
	g_{cm} 		&	g_{dm}	&	\omega_m -\frac{i\gamma_m}{2} 	&	g_{mn}		\\
	g_{cn}		&	g_{dn}	&	g_{mn}	&	\omega_ n -\frac{i\gamma_n}{2} 	\\
	\end{pmatrix},
\label{eq:mmc_osc}
\end{equation}
where $g$, $\omega$ and $\gamma$ indicate their couplings, resonant frequencies and dissipations, respectively.
Fig.\,\ref{fig:tuning} (right) shows the function $f_{cdmn}(\omega,\omega_m) = \det \big( \omega\mathbb{I}_4 - {\cal H}_{cdmn} \big)$, whose maxima identify the resonance frequencies of the HS. 
By comparing the two plots of Fig.\,\ref{fig:tuning}, one can see that the model appropriately describes the system, allowing us to extract the linewidths, frequencies and couplings of the modes through a fit.
The typical measured values are $\gamma_1\simeq1.9\,$MHz and $g_{cm}\simeq638\,$MHz, yielding $\tau_s\simeq84\,$ns and $N_s\simeq1.0\times10^{21}\,$spins, respectively.
Remarkably, the mode $\omega_1$ is not altered by other modes, thus we will use it to search for axion-induced signals.
For a fixed $B_0$ the linewidth of the hybrid mode is the haloscope sensitive band. By changing $B_0$, we can perform a frequency scan along the dashed line of Fig.\,\ref{fig:tuning}. 

	\begin{figure}[h!]
	\centering
	\includegraphics[width=.475\textwidth]{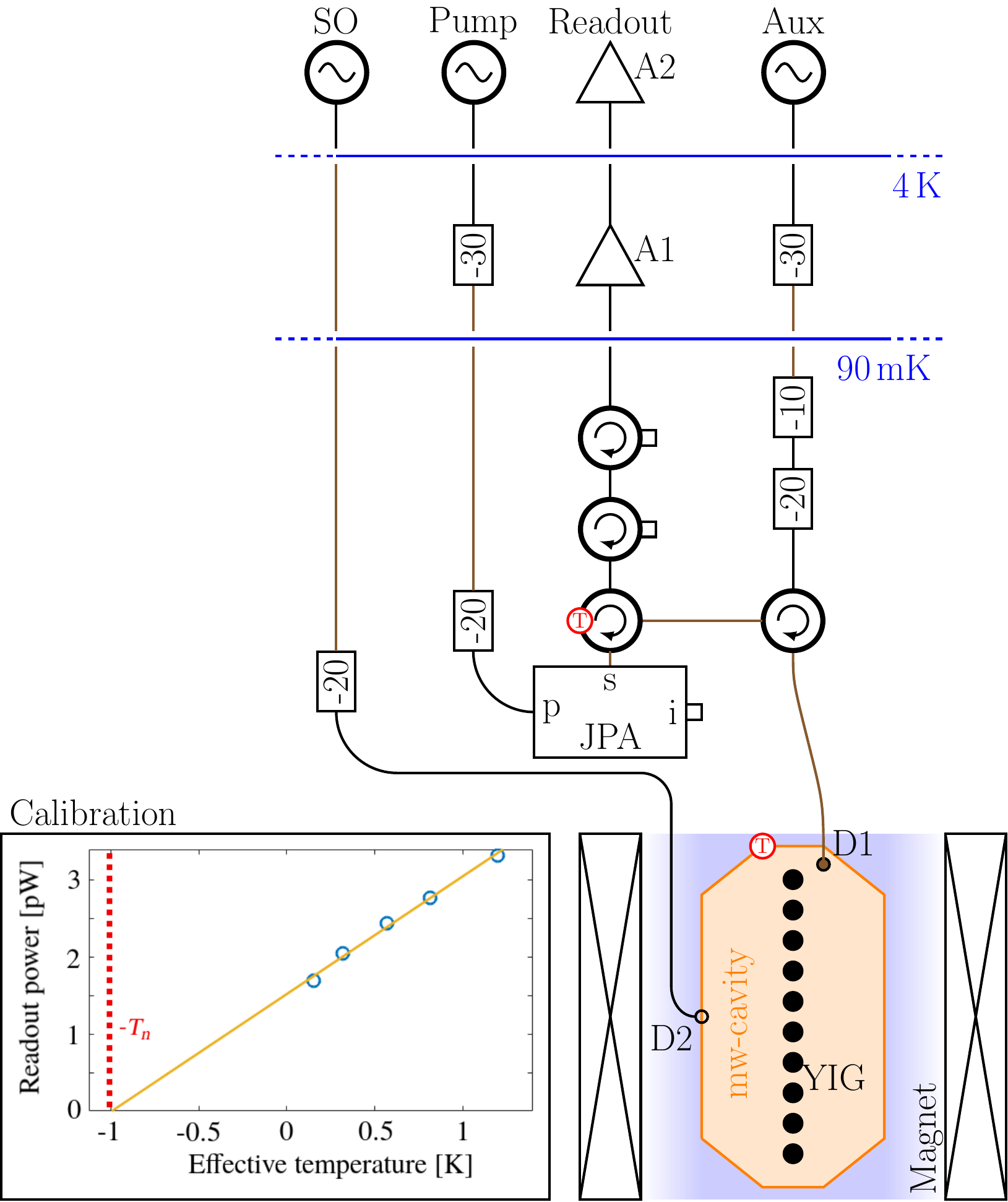}
	\caption{Schematics of the apparatus. The cavity is reported in orange, the ten YIG spheres are in black, and the blue shaded region is permeated by a uniform magnetic field. The cryogenic and room temperature HEMT amplifiers are A1 and A2, respectively, and the JPA ports are the signal (s), idler (i) and pump (p). Superconducting cables are brown, the red-circled $T$s are the thermometers, SO is a source oscillator, and attenuators are shown with their reduction factor in dB. As inset, we show the calibration of the system gain and noise temperature, obtained by injecting signals in the SO line. The power injected in the HS is given in terms of an effective temperature proportional to $A_\mathrm{cal}$. The errors are within the symbol dimension. See text for further details.}
	\label{fig:app}
	\end{figure}	
The electronic schematics, shown in Fig.\,\ref{fig:app}, consists in four rf lines used to characterize, calibrate and operate the haloscope. 
The HS output power is collected by a dipole antenna (D1), connected to a manipulator by a thin steel wire and a system of pulleys to change its coupling.
The source oscillator (SO) line is connected to a weakly coupled antenna (D2) and used to inject signals into the HS, the Pump line goes to a Josephson parametric amplifier (JPA), the Readout line amplifies the power collected by D1, and Aux is an auxiliary line. 
The Readout line is connected to an heterodyne as described in \cite{quaxepjc}, where an ADC samples the down-converted power which is then stored for analysis.
The JPA is a quantum limited amplifier, with resonance frequency of about 10\,GHz resulting in a noise temperature of 0.5\,K. Its gain is close to 20\,dB in a band of order 10\,MHz, and its working frequency can be tuned thanks to a small superconducting coil \cite{PhysRevLett.108.147701}. Excluding some mode crossings, hybrid mode and JPA frequencies overlap between 10.2\,GHz and 10.4\,GHz, and allow us to scan the corresponding axion mass range. 

The procedure to calibrate all the lines of the setup is:
(\textit{i}) the transmittivity of the Aux-Readout path $K_\mathrm{AR}$ is measured by decoupling D1 or by detuning $\omega_1$;
(\textit{ii}) for the Aux-SO and SO-Readout paths, $K_\mathrm{AS}$ and $K_\mathrm{SR}$ are obtained by critically coupling D1 to the mode $\omega_1$.
The transmittivity of the SO line is $K_\mathrm{SO}\simeq\sqrt{K_\mathrm{SR}K_\mathrm{AS}/K_\mathrm{AR}}$.
If a signal of power $A_\mathrm{in}$ is injected in the SO line, the fraction of this power getting into the HS results $A_\mathrm{cal}=A_\mathrm{in}K_\mathrm{SO}$.
Since $A_\mathrm{cal}$ is a calibrated signal, it can be used to measure gain and noise temperature of the Readout line. From this measurement we obtain a system noise temperature $T_n=1.0\,$K, and a gain of 70.4\,dB from D1 to Readout (see Fig.\,\ref{fig:app}).
In our setup, the coupling of D1 can be varied of 8\,dB, thus we estimate a calibration uncertainty of 16\%.
We measured the JPA gain, the HEMTs noise temperature, and the cavity temperature to get the noise budget detailed in Tab.\,\ref{tab:noisebudget}. The 0.12\,K difference may be due to unaccounted losses, or non-precise temperature control.	
	\begin{table}[h!]
	\begin{tabular}{c | c  }
		Source	&	Estimated	 \\
	\hline
	Quantum noise	&	0.50\,K	\\
	Thermal noise	&	0.12\,K	\\
	HEMTs noise		&	0.25\,K	\\
	\hline
	\hline
					&			\\
	Expected total	&	0.87\,K	\\
	\hline
	& \\
	Measured total $T_n$		&	0.99\,K 	\\
	\end{tabular}
	\caption{Noise budget of the apparatus. The measured noise is compatible with the estimated one.}
	\label{tab:noisebudget}
	\end{table}
	
	\begin{figure}[h!]
	\centering
	\includegraphics[width=.45\textwidth]{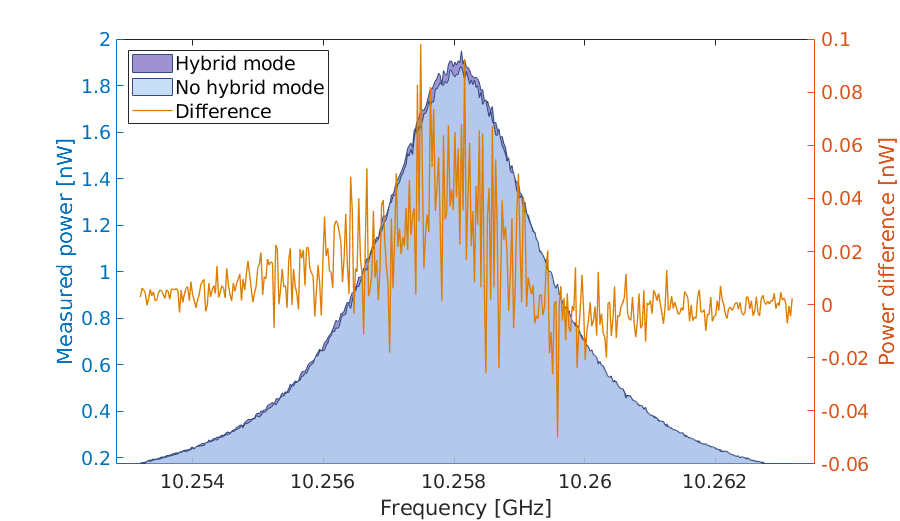}
	\caption{Thermal noise of the HS. The blue curves are the power measured at the Readout with $\omega_1$ in the JPA bandwidth (dark blue) and out of it (light blue). The difference between the two is the HS noise (reported in orange). 
	}
	\label{fig:thermal}
	\end{figure}	
To double check the accuracy of the result, we measure the thermal noise of the HS. 
The noise difference for $\omega_1$ on and off the JPA resonance (dark blue and light blue) gives the noise added by the hybrid mode (orange curve), as shown in Fig.\,\ref{fig:thermal}.
The excess noise is compatible with a temperature of the HS $\sim10\,$mK higher than the one of the nearest load, which is realistic. Similar results are obtained by changing the D1 antenna coupling for a fixed $B_0$.

The axion search consisted in fifty-six runs, each one with fixed $B_0$.
For every run a transmission measurement of the hybrid system is used to set $\omega_1$, to critically couple D1 to it, and to measure $\gamma_1$. The frequency stability of $\omega_1$ resulted well below the linewidth within an interval of several hours, allowing long integration times. 
Data are stored with the ADC over a 2\,MHz band around $\omega_1$ for subsequent analysis.
We FFT the data with a 100\,Hz resolution bandwidth to identify and remove biased bins and disturbances in the down-converted spectra.
To estimate the sensitivity to the axion field, we rebin the FFTs with a resolution $\mathrm{RBW}\simeq5\,$kHz, which at this frequency gives the best SNR for the axionic signal \cite{BARBIERI1989357}. The spectra are fitted to a degree five polynomial to extract the residuals, whose standard deviation is the sensitivity of the apparatus. We verified that the analysis procedure excludes unwanted bins while preserving the signal and SNR by adding a fake axion signal to real data.

	\begin{figure*}
	\centering
	\includegraphics[width=.9\textwidth]{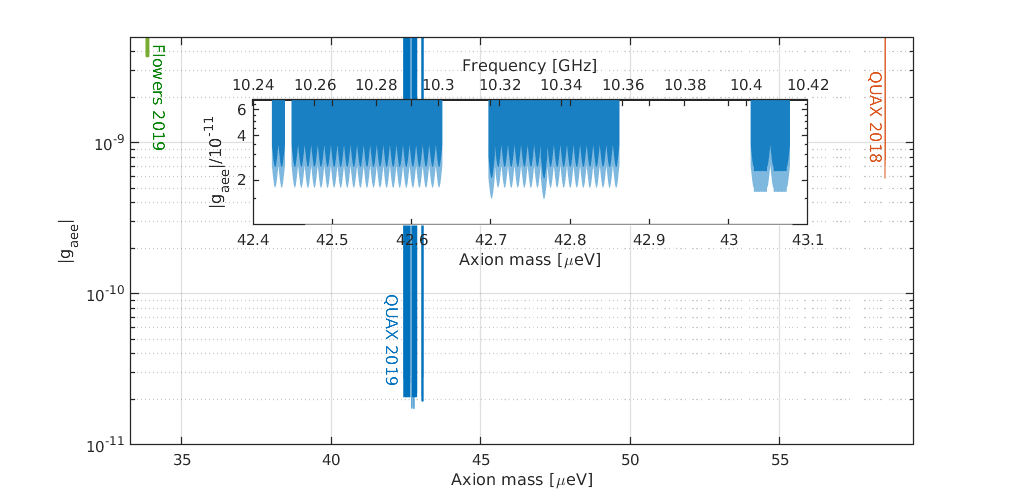}
	\caption{Exclusion plot at 95\% CL on the axion-electron coupling obtained with the present prototype (excluded region reported in blue and error in light blue), and overview of other searches for the axion-electron interaction. The other results are from \cite{quaxepjc} (orange) and \cite{FLOWER2019100306} (green), while the DFSZ axion line is at about $g_{aee}\simeq10^{-15}$. The inset is a detailed view of the reported result.}
	\label{fig:limit}
	\end{figure*}
	
Our data were collected in July 2019 in a total run time of 74\,h. The average run length was $\sim1\,$h, and each one was performed during the maximum of the daily-modulated axionic signal. 
The measured fluctuations are compatible with the estimated noise in every run, and we detected no statistically significant signal consistent with the DM axion field. The minimum measured fluctuation is $\sigma_P=5.1\times10^{-24}\,$W, for the longest integration time $t=9\,$h, where the Dicke prediction is $k_B T_n\sqrt{\mathrm{RBW}/t}=4.8\times10^{-24}\,$W.
In terms of rf magnetic field, this result corresponds to $\sigma_B=5.5\times10^{-19}\,$T, which, to our knowledge, is a record one for an rf spin-magnetometer.
The absence of fast rf bursts in the data is verified by using a 1\,ms time resolution waterfall spectrogram.

Even if the minimum field detectable by the haloscope is much larger than $B_a$, these measurements can still be a probe for ALPs, which may also constitute the totality of DM \cite{Arias_2012}.	
The 95\% CL upper limit on the axion electron coupling constant is
	\begin{equation}
	g_{aee}<\frac{e}{\pi m_av_a}\sqrt{\frac{k_{ac}\times2\sigma_P}{2\mu_B \gamma_e\, n_a N_s \tau_s}}\simeq1.7\times10^{-11}.
	\label{eq:gaee}
	\end{equation}
The transduction coefficient of the axionic signal $k_{ac}$ was calculated with a model similar to the one of Eq.\,(\ref{eq:mmc_osc}) \cite{rfband}. It essentially depends on $\omega_1$ and, in our bandwidth, results $0.5<k_{ac}<1.0$.
The overall exclusion plot obtained with the ferromagnetic haloscope is given in Fig.\,\ref{fig:limit}. 
All the experimental parameters used to extract the limits from Eq.\,(\ref{eq:gaee}) are measured within every run, making the measurement highly self-consistent.

These results improved the best previous limits \cite{quaxepjc} by roughly a factor 30 in $g_{aee}$ and 50 in bandwidth.
The improvement over the previous prototype is due to an increased material volume, to an almost quantum-limited noise temperature, and to longer integration times. No axion-mass scan was performed by previous experiments of this kind, and we now demonstrate that it is feasible to tune a hybrid resonance over hundreds of MHz to search for axion-deposited power. Our prototype scanned a range of axion masses of about 0.7\,$\mu$eV with a field variation of 7\,mT, drastically simplifying the tuning of the haloscope.

In conclusion, we designed and developed a quantum-limited rf spin-magnetometer used as an axion haloscope. 
The instrument implements an axion-to-rf transducer, i.\,e. an hybrid system which embeds one of the largest quantity of magnetic material to date, and a detection electronics based on a quantum-limited JPA.
The operation of this instrument led to an axion search over a span of $0.7\,\mu$eV around $42.7\,\mu$eV, with a maximum sensitivity to $g_{aee}$ of $1.7\times10^{-11}$. This, to our knowledge, is the best reported limit on the coupling of DM axions to electrons, and corresponds to a 1-$\sigma$ field sensitivity of $5.5\times10^{-19}\,$T, which is a record one.
No showstoppers have been found so far, and hence a further upscale of the system can be foreseen. 
A superconducting cavity with a higher quality factor was already developed and tested \cite{PhysRevD.99.101101}. It was not employed in this work since the YIG linewidth does not match the superconducting cavity one, and the improvement on the setup would have been negligible.
With this prototype we reached the rf sensitivity limit of linear amplifiers \cite{RevModPhys.82.1155}. To further improve the present setup one needs to rely on bolometers or single photon/magnon counters \cite{PhysRevD.88.035020}. Such devices are currently being studied by a number of groups, as they find important applications in the field of quantum information \cite{PhysRevLett.117.030802,Inomata2016,8395076,lachancequirion2019entanglementbased}.

We are grateful to E. Berto, A. Benato and M. Rebeschini, who did the mechanical work, F. Calaon and M. Tessaro who helped with the electronics and cryogenics, and  to F. Stivanello for the chemical treatments. We thank G. Galet and L. Castellani for the development of the magnet power supply, and M. Zago who realized the technical drawings of the system. We deeply acknowledge the Cryogenic Service of the Laboratori Nazionali di Legnaro, for providing us large quantities of liquid helium on demand.

\bibliographystyle{unsrt}
\bibliography{quax_gaee19}

\end{document}